\documentstyle[sphonnef,pslatex]{article}  
\newcommand{\plotinsert}[4]{\vspace*{#1}
\begin{center}\leavevmode\epsfxsize=#2
\epsfbox{#3}\end{center}\vspace*{#4}} 
\newcommand{\mycaption}[2]{\caption[]{#1}\vspace*{#2}}
\frompage{000} \topage{000}                                              
\def\rightmark{Hadron Structure Studied with the Electromagnetic Probe} 
\def\leftmark{H. Arenh\"ovel}
\title{
Hadron Structure Studied with the Electromagnetic Probe -- from 
Giant Resonances to Meson Production
} 
\authors{
{Hartmuth Arenh\"ovel%
}\\[2.812mm]
{\normalsize
Institut f\"ur Kernphysik, Johannes Gutenberg-Universit\"at, \\ 
D-55099 Mainz, Germany\\[0.2ex] 
}}
 
\abstract{The development of theoretical photonuclear physics over the 
last 40 years is illustrated by a few selected examples highlighting a 
number of important issues like collective motion in nuclei, the role of 
subnuclear degrees of freedom, relativity and meson production.}
 
\begin{document}

\maketitle
\vspace*{24pt}

\section{Introduction}

In this talk, which I dedicate to the memory of Michael Danos, 
I would like to give a brief overview on a number of important issues 
in the field of photonuclear physics which have deepend considerably our 
understanding of nuclear structure over the past 40 years. It is not possible 
to give here a comprehensive review, and I had to make a selection which 
is governed by the scientific work of Danos, who has made many important 
contributions to nuclear photoreactions, and also 
by my own research interests. A very good represention of benchmark 
papers in this field may be found in~\cite{FuH76}. Let me remind you
that in nuclear physics the electromagnetic probe has always been a very 
important tool. This is illustrated by a few highlights 
from the early development of nuclear physics:
\begin{itemize}
\item
1909: Coulomb scattering of $\alpha$-particles from a gold foil by 
Geiger and Marsden~\cite{GeM09} from which Rutherford concluded in 
1911 that most of the mass of an atom is concentrated in a tiny, almost 
pointlike nucleus in its center~\cite{Rut11}.
\item
1934: Photodisintegration of the deuteron by Chadwick and 
Goldhaber~\cite{ChG34} marks the beginning of photonuclear physics.
\item
1947: Discovery of the giant dipole resonance by Baldwin and 
Klaiber~\cite{BaK47} as a collective phenomenon almost exhausting the 
Thomas-Reiche-Kuhn dipole sum rule completely.
\item
1951: First electron scattering experiment by Lyman, Hanson and 
Scott~\cite{LyH51} establishing electron scattering as a very important tool 
for nuclear structure studies.
\end{itemize}
The important role of Danos's research in the field of photonuclear 
physics is best illustrated by his influential 1961-Lectures at the 
University of Maryland~\cite{Dan61}. In these lectures Danos covered 
quite a variety of different topics: properties 
of the e.m.\ field, multipole decomposition, quantization, gauge invariance, 
Siegert's theorem, dispersion relations, TRK and other sum rules, to name a 
few.

\section{The dynamic collective model of the giant resonances}

The collective phenomenon of the giant dipole resonance (GDR) can well be 
explained in the framework of the hydrodynamical model of Steinwedel 
and Jensen as an oscillation of a proton fluid against a neutron fluid. 
In particular, the dependence of the position of the GDR on the mass number 
is governed by the fact that for a spherical nucleus the eigenfrequency is 
proportional to the nuclear radius. It was the important observation of 
Danos (and independently of Okamoto) that for an axially symmetric 
deformed nucleus the GDR will be split into two peaks corresponding 
to oscillations along and perpendicular to the symmetry axis~\cite{Dan58}. 

\begin{figure}[h]
 \plotinsert{-.5cm}{9cm}{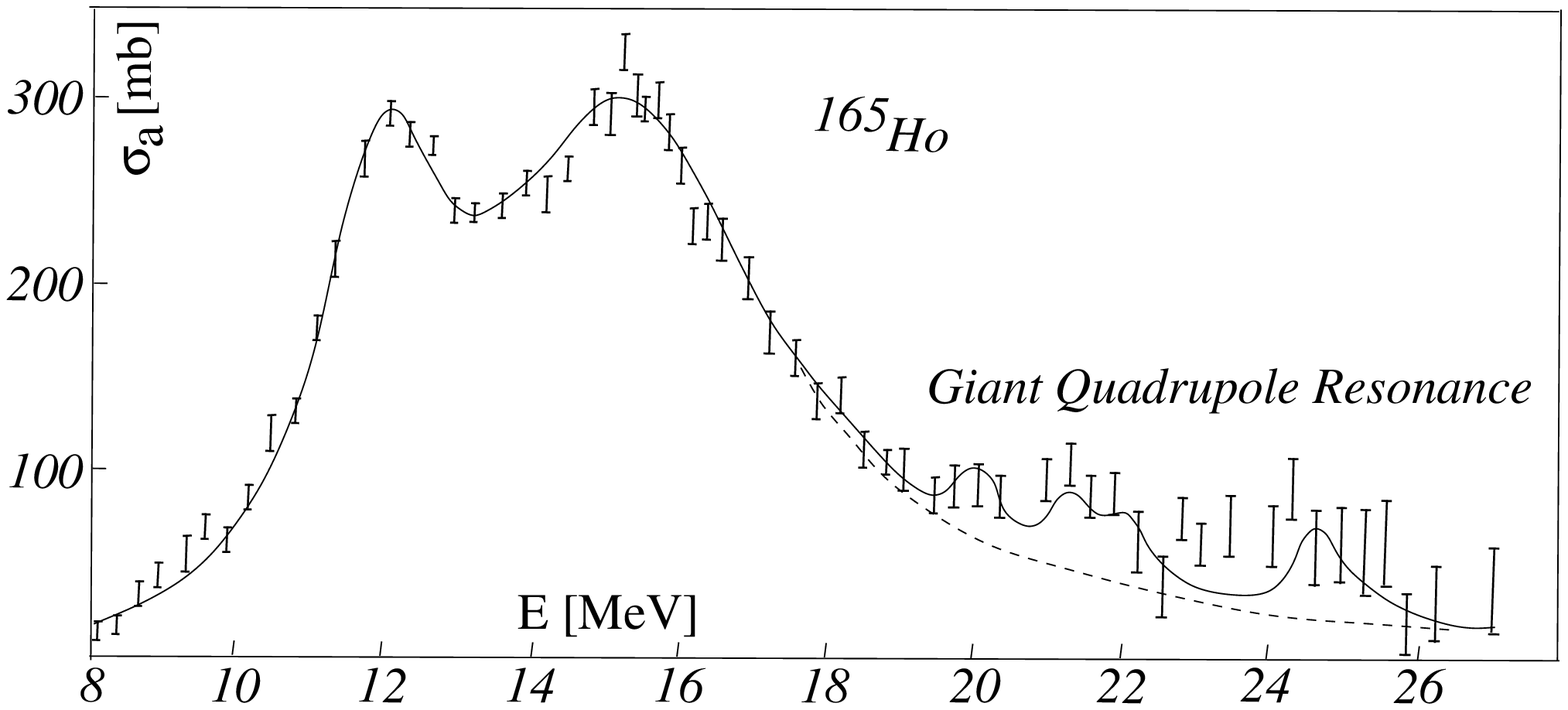}{-.5cm}
\mycaption{
E1 and E2 photoabsorption cross section for 
$^{165}$Ho (from~\cite{LiG66}). Experimental data from \cite{Bra63}.
}{-.3cm}
\label{fig1}
\end{figure}

In view of the additional collective surface degrees of freedom, Danos and 
Greiner~\cite{DaG64} developed in 1964 a unified dynamic collective model of 
the giant resonances (DCM) which includes in particular the coupling between 
the rotation-vibration and the giant resonance d.o.f.\ 
leading to additional dynamic effects. A comparison between the predictions 
for the total $\gamma$-absorption cross section and experimental data is 
shown in Fig.~\ref{fig1} for $^{165}$Ho including the giant quadrupole 
resonance contributions. The very good agreement is evident. 

\begin{figure}[h!]
\plotinsert{-.3cm}{12cm}{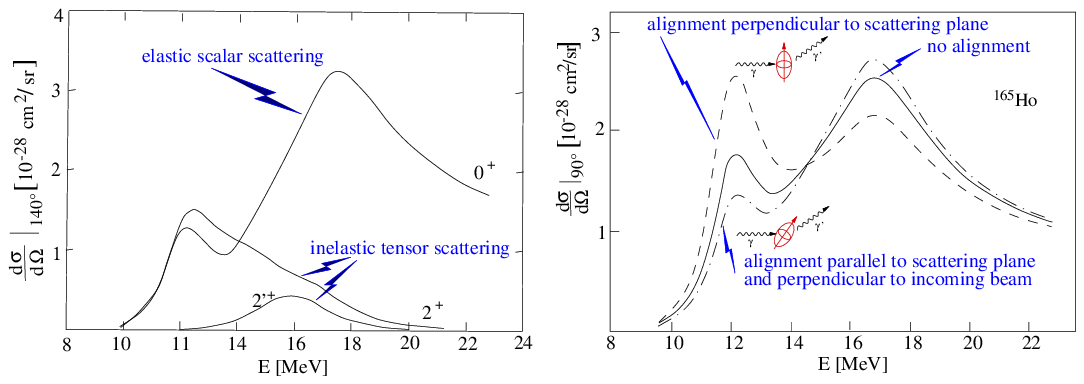}{-.6cm}
\mycaption{
Left panel: Calculated elastic and inelastic photon scattering cross 
sections at $140^\circ$ for $^{166}$Er (from~\cite{ArG67}). 
Right panel: Elastic photon scattering cross sections for $^{165}$Ho 
(from~\cite{ArG66}):
unoriented target: solid curve, aligned target:
(a) perpendicular to scattering plane: dashed curve,
(b) parallel to scattering plane and perpendicular to incoming
photon beam: dash-dot curve.
}{-.2cm}
\label{fig2}
\end{figure}

An important consequence of this dynamic coupling is the considerably strong 
dipole transition strength from the GDR states to the low lying rotational and 
vibrational states leading to sizeable Raman scalar and tensor scattering 
into these states (see Fig.~\ref{fig2}, left panel) as experimentally 
observed by Fuller and Hayward~\cite{FuH76}. Another interesting feature of 
the DCM is the fact that a deformed nucleus with a
nonvanishing ground state spin becomes optically anisotropic (nonvanishing 
tensor polarization) and thus its absorption and scattering cross sections 
depend on the nuclear orientation. This is shown in the right panel of 
Fig.~\ref{fig2} for $^{165}$Ho with spin $I=7/2$. A review may be found 
in~\cite{ArG69}.
\begin{figure}[h!]
  \plotinsert{-.2cm}{7cm}{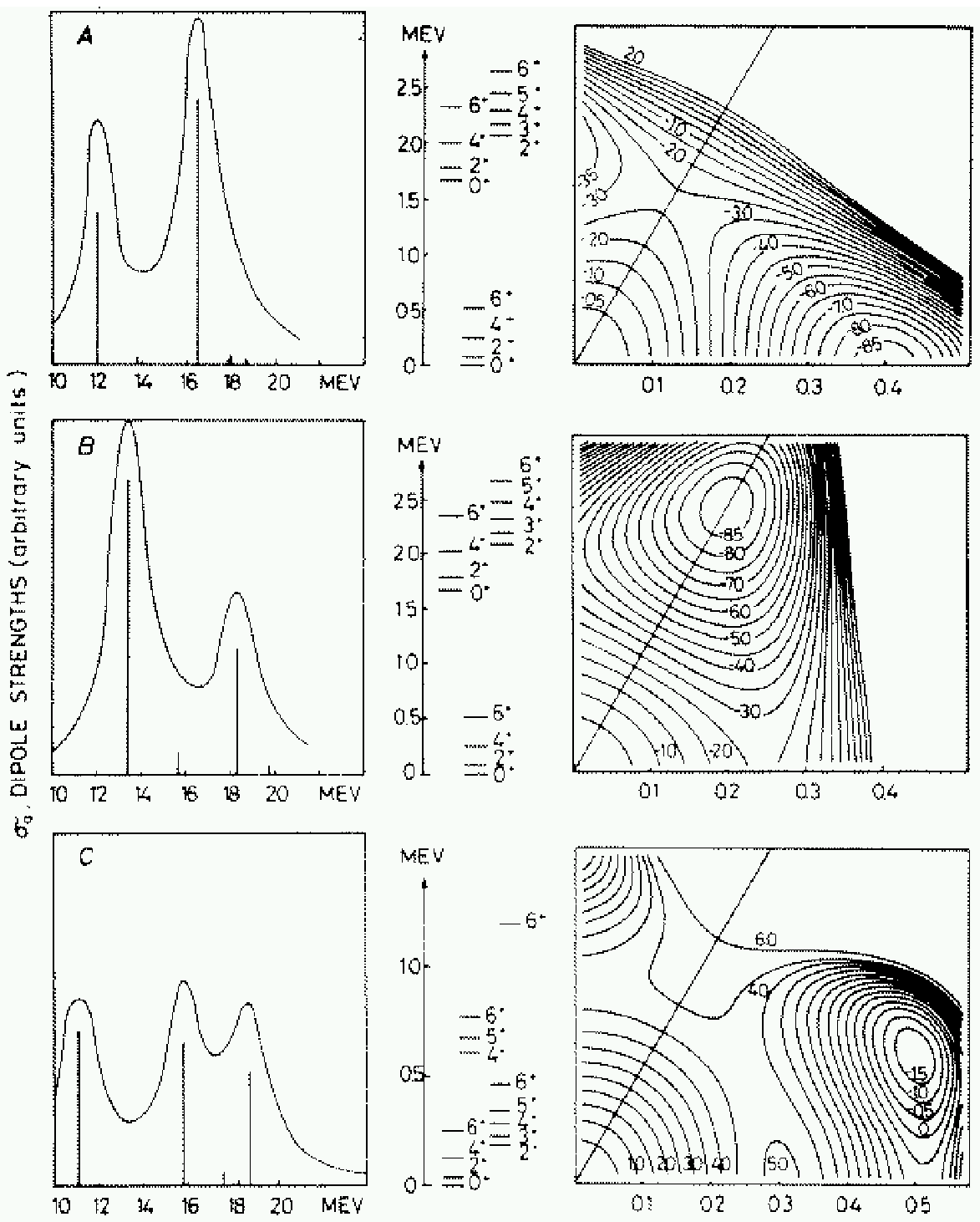}{-.6cm}
\mycaption{
Dipole strengths and photon absorption cross sections, low energy 
spectra and potential energy surfaces for an axially symmetric  
prolate (A), oblate (B), and triaxially deformed nucleus (from~\cite{ReG70}).
}{-.4cm}
\label{fig4}
\end{figure}
Subsequently, the DCM was further developed in order to describe a much more 
general class of potential energy surfaces for the low energy collective 
d.o.f.\ allowing a unified description of the GDR for nuclei with quite 
different collective characteristics~\cite{ReG72}. An example is shown in 
Fig.~\ref{fig4}.

\section{Subnuclear degrees of freedom}
\begin{figure}[h!]
  \plotinsert{-.5cm}{10cm}{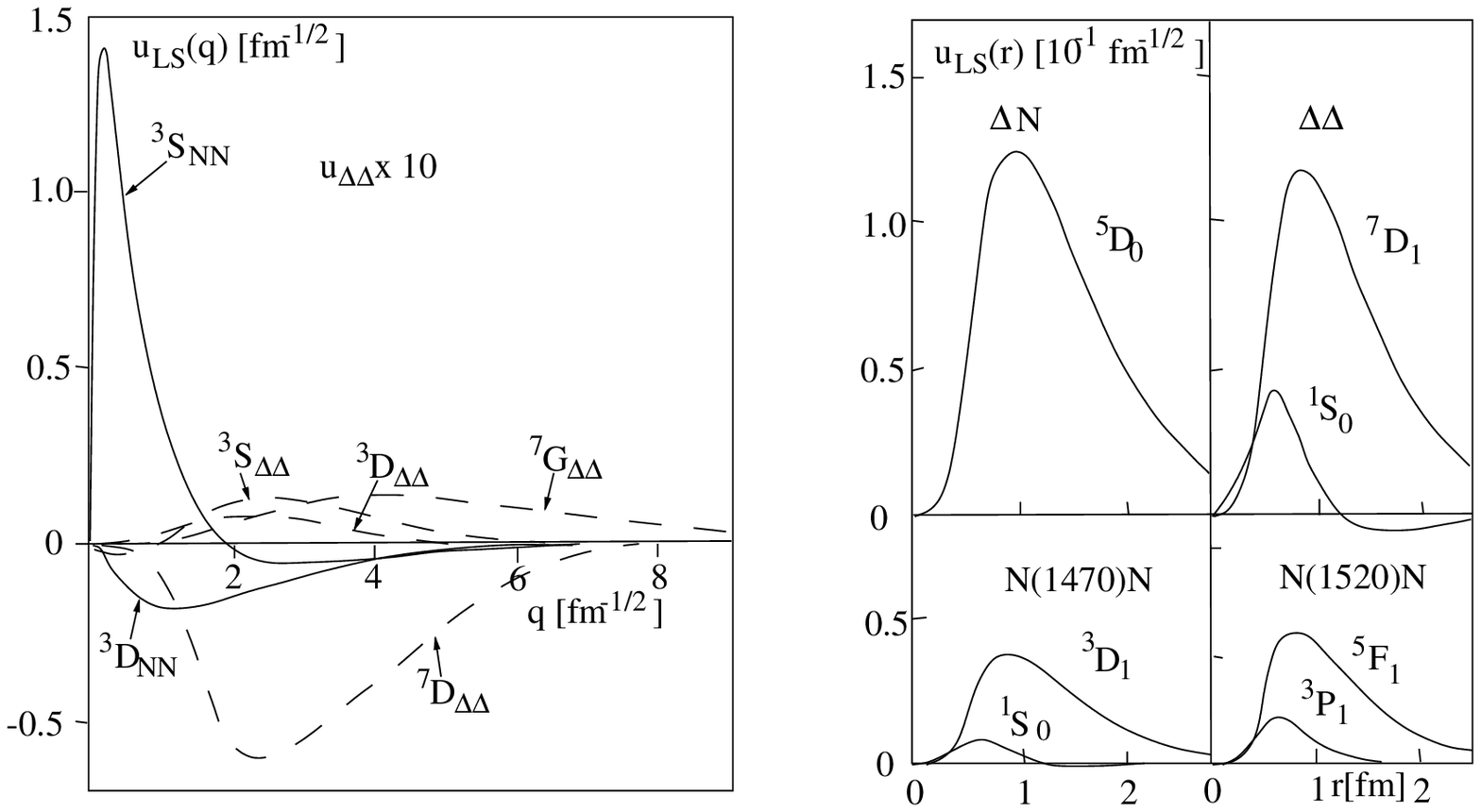}{-.7cm}
\mycaption{Left panel: 
Normal and $\Delta\Delta$-component of 
the deuteron ($P_{\Delta\Delta}=.97$~\%) (from~\cite{Are75}).
Right panel: Relative two-particle wave 
functions of IC in $^4$He (from~\cite{HoA78}). 
}{-.7cm}
\label{fig5}
\end{figure}

In his 1961-Lectures Danos had already alluded to the possibility that a 
nucleon, in view of its extended internal structure, could be slightly 
deformed in a nuclear medium. Indeed, at the end of the 60s several groups, 
including Danos, H.T.\ Williams and myself, 
were developing a phenomenological model of such modifications by admixing 
into the nuclear wave function configurations, where one or several nucleons 
are internally excited, say as a $\Delta(1232)$-resonance~\cite{ArD70}. 
Such wave function components have been coined 
``nuclear isobar configurations'' (IC). Reviews may be found 
in~\cite{Gre76,WeA78}. The characteristic features of these 
IC is that they possess small probabilities and a short range structure. 
Examples are shown in Fig.~\ref{fig5}. 

\begin{figure}[h!]
 \plotinsert{-.2cm}{10cm}{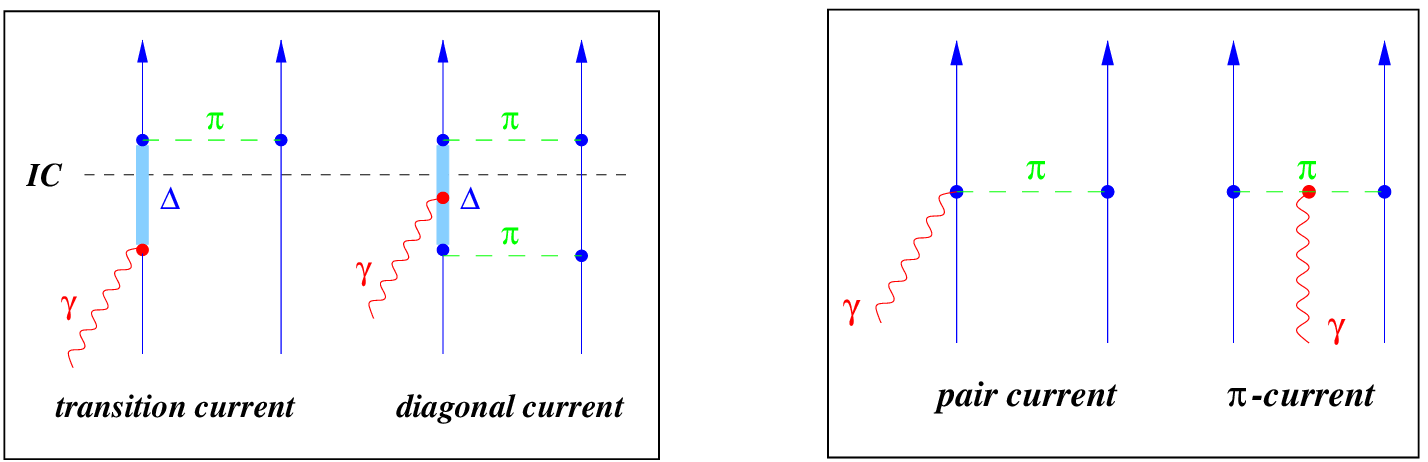}{-.7cm}
\mycaption{Left panel: Effective MEC operator with intermediate $N\Delta$ 
configuration; Right Panel: Pion exchange current.
}{-.2cm}
\label{fig6}
\end{figure}

\begin{figure}[h!]
  \plotinsert{-.4cm}{12cm}{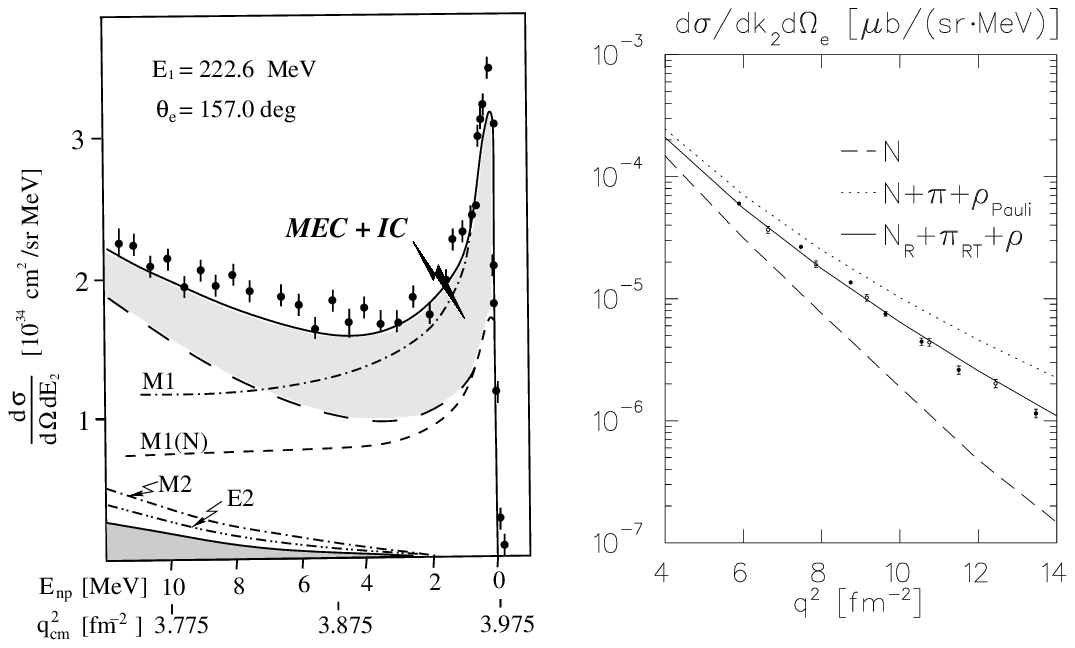}{-.5cm}
\mycaption{Left panel: Inclusive cross section for $d(e,e')$ 
as function of energy $E_{np}$ (from~\cite{FaA76}); 
Right Panel: Inclusive cross section for $d(e,e')$, averaged over 
$E_{np}=0-3$ MeV, as function of momentum transfer $q^2$ (from~\cite{Rit95}).
}{-.5cm}
\label{fig7}
\end{figure}

These subnuclear d.o.f., as manifest in isobar admixtures in nuclear wave 
functions, contribute also to the e.m.\ interaction. Alternatively, 
their contribution may be described by effective two-body exchange currents 
involving IC 
as shown diagrammatically in the left panel of Fig.~\ref{fig6} together 
with similar current contributions (MEC) from meson d.o.f.\ mediating the 
strong interaction, shown in the right panel for $\pi$-exchange. The largest 
MEC effects are found in $E1$-transitions which, however, are largely covered 
by the Siegert operator~\cite{Are81}. Particularly strong MEC contributions 
to $M1$ are found in deuteron electrodisintegration near threshold, 
for which I show the inclusive cross section in Fig.~\ref{fig7}. One readily 
notices the sizeable increase of the cross section by MEC and IC, particularly 
dominant at higher momentum transfers (see also~\cite{HoR73,LoF75}).

\begin{figure}[h!]
  \plotinsert{-.3cm}{12cm}{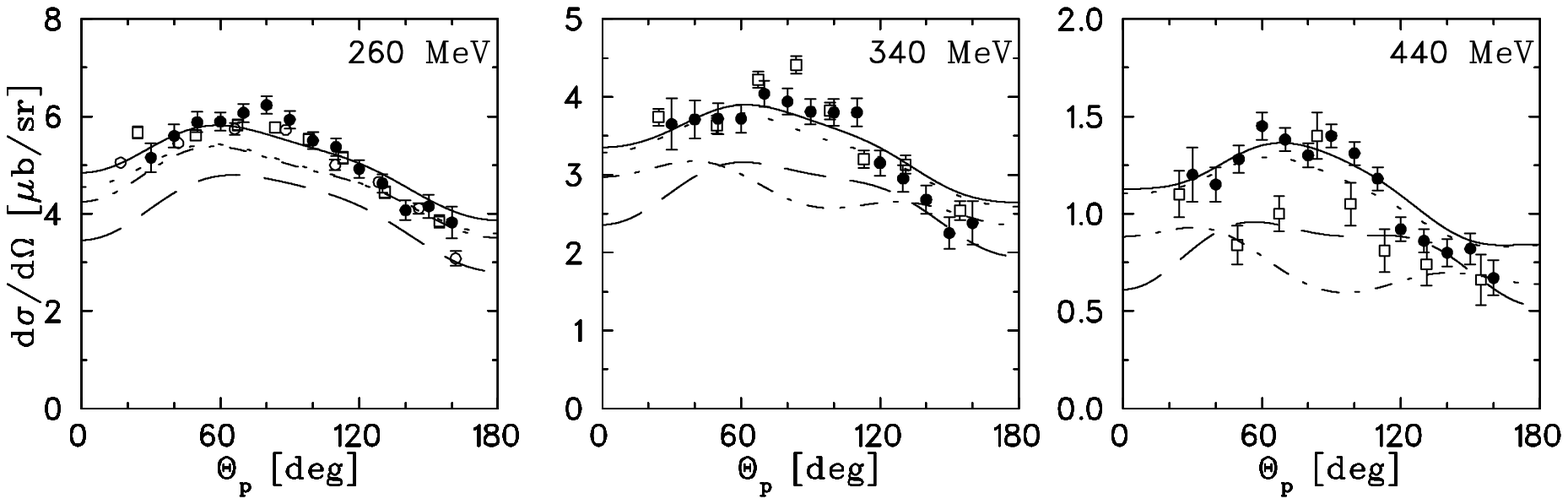}{-.7cm}
\mycaption{
Differential cross section for $d(\gamma,p)n$ (from~\cite{ScA00}): 
dash-dot: static interaction 
and MEC; dashed: retarded interaction, static MEC; solid: 
retarded interaction and MEC; dotted: as full curves, but no $\pi d$-channel.
Exp: 
$\bullet$ J. Arends {\it et al.}, Nucl. Phys. A {\bf 412}, 509 (1984), 
$\Box$  G. Blanpied {\it et al.}, Phys. Rev. C {\bf 52}, R455 (1995),
$\circ$  R. Crawford {\it et al.}, Nucl. Phys. A {\bf 603}, 303 (1996).
}{-.3cm}
\label{fig8}
\end{figure}

As last example for the importance of subnuclear d.o.f.\ I show in 
Fig.~\ref{fig8} differential cross sections of deuteron photodisintegration 
in the $\Delta$-resonance region, where $\Delta$-d.o.f.\ dominate. For a long 
time, the most sophisticated theoretical approaches were unable to describe 
properly the experimental date in this energy region. Only recently, a careful 
analysis has shown, that a proper treatment of meson retardation is absolutely 
necessary for a satisfactory description of experimental data~\cite{ScA00}.  

\section{Relativistic effects in the GDH sum rule}

The Gerasimov-Drell-Hearn sum rule (1965/66) links a ground state 
property, the anomalous magnetic moment, to the energy weighted integral 
from threshold up to infinity over the spin asymmetry 
$\sigma ^P(k)-\sigma ^A(k)$, the difference of the total 
photoabsorption cross sections for circularly polarized photons on a target 
with spin parallel and antiparallel to the spin of the photon, i.e.
\begin{eqnarray*}
\int_0^\infty \frac{dk}{k}\Big(\sigma^P(k)-\sigma^A(k)\Big)
\,&=&\,\, 4\,\pi^2 {\kappa^2}\frac{e^2}{M_t^2}\,I\,,
\end{eqnarray*}
where $I$ denotes the spin of the particle, $M_t$ its mass, and
${\kappa}$ its anomalous magnetic moment, defined by the 
total magnetic moment $\vec M = (Q+ \kappa)\frac{e}{M_t}{\vec S}$,
with $eQ$ as its charge and $\vec S$ its spin operator. 
An important consequence is, that a particle has 
to possess an internal structure if $\kappa \neq 0$. 

Here I will consider the GDH sum rule and in particular the spin asymmetry 
for the deuteron~\cite{ArK97} 
in order to illustrate the fact, that even at low energies relativistic 
effects can become quite important in certain polarization observables.
Application of the GDH sum rule to the deuteron reveals a very interesting 
feature. On the one hand, one expects a very small anomalous magnetic moment 
for the deuteron, because the deuteron has isospin zero ruling out the 
contribution of the large nucleon isovector anomalous magnetic moments to 
the deuteron magnetic moment. In fact, the experimental value 
is $\kappa_d=-.143$ resulting in a GDH prediction of $I^{GDH}_d = 0.65\,\mu$b, 
which is more than two orders of magnitude smaller 
than the nucleon values. 
On the other hand, one has the following absorptive processes: (i) 
photodisintegration $\gamma + d \, \rightarrow  \,n + p$,
(ii) single pion production 
(coherent: $\gamma + d \, \rightarrow  \,d + \pi^0$, and 
incoherent: $\gamma + d \,\rightarrow  \,N+N+\pi $),
(iii) two pion production etc. 
The processes (ii) and (iii) are dominated by quasifree production and,
therefore, one may estimate from them a positive GDH contribution of the 
order of the sum of proton and neutron, namely 
$I^{GDH}_p(\infty)+ I^{GDH}_n(\infty)=438\, \mu\mbox{b}$. 
In order to obtain the small total 
deuteron GDH value one, therefore, needs a large negative contribution of 
about the same size for cancellation. 

\begin{figure}[h]
 \plotinsert{-.4cm}{10cm}{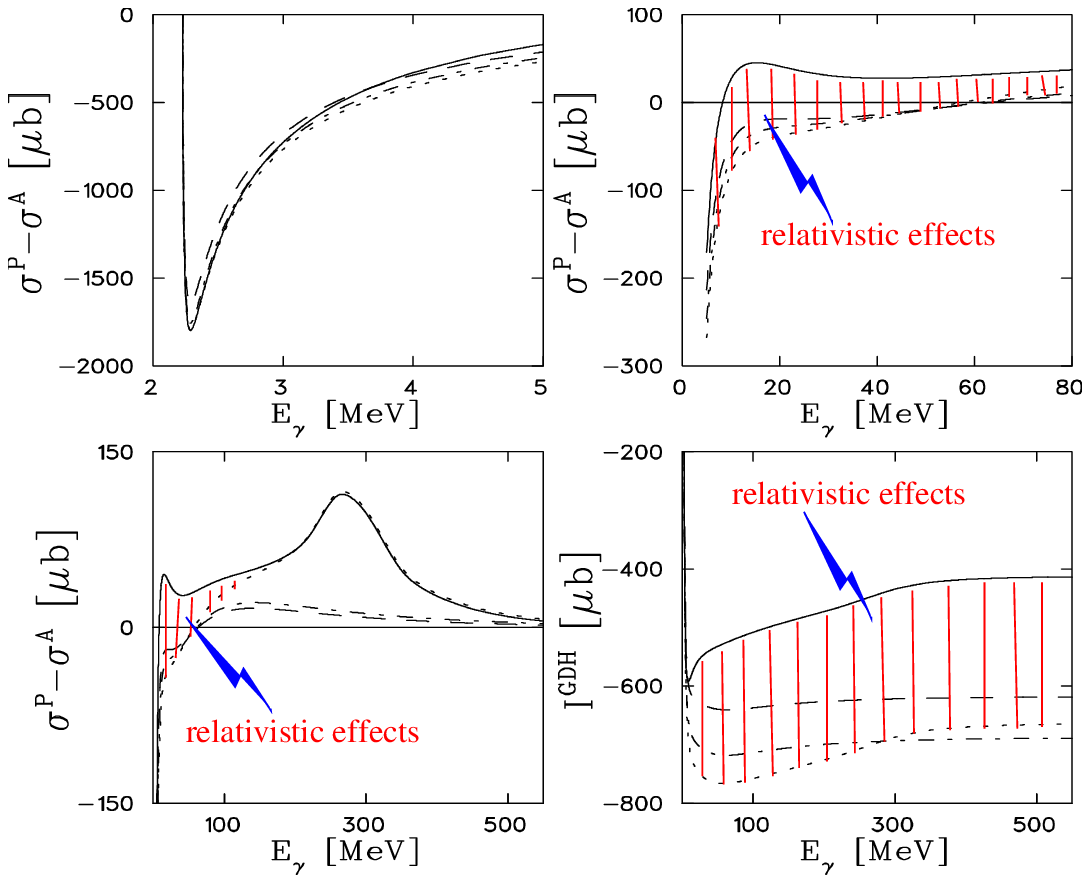}{-.7cm}
\mycaption{Spin asymmetry and integrated GDH-integral for deuteron 
photodisintegration (from~\cite{ArK97}). 
Two upper and lower left panels: difference of the cross sections 
in various energy regions; lower right panel: 
$I^{GDH}_{\gamma d \to np}$ as function of the upper integration energy. 
dashed: N (normal, including Siegert MEC); 
dash-dot: N + MEC; dotted: N + MEC + IC; solid: total (N + MEC + IC + {RC}).
}{-.5cm}
\label{fig9}
\end{figure}

Indeed, from the photodisintegration channel, which is 
the only photoabsorption process below the pion production threshold, a 
sizeable negative contribution arises at very low energies near threshold 
from the $M1$-transition to the 
resonant $^1S_0$ state, because this state can only be reached if the spins 
of photon and deuteron are antiparallel, and is forbidden for the parallel 
situation. The photodisintegration channel has been evaluated within 
a nonrelativistic framework but with inclusion of the most important 
relativistic contributions. The results are summarized in Fig.~\ref{fig9}, 
where the spin asymmetry and the GDH integral as function of the upper 
integration limit is shown. One readily notices a very drastic decrease of 
the spin asymmetry at quite low energies by relativistic contributions, here 
mostly by the leading order relativistic spin-orbit current. 
At about 500 MeV convergence of the GDH-integral is achieved for the 
photodisintegration channel yielding a sume rule contribution of $-413$ $\mu$b
cancelling almost completely the estimated contribution from meson production. 
Without the relativistic contributions, one would 
have found a GDH contribution from
photodisintegration of $-619$ $\mu$b, which would lead to a negative total 
GDH-value for the deuteron. This demonstrates very convincingly the importance 
of relativistic effects in the spin asymmetry.

\section{Eta photoproduction on the deuteron}

Finally, I will briefly consider photoproductions of mesons on the deuteron. 
Meson photoproduction is the primary absorptive process on the nucleon 
and the interest of meson production on nuclei arises from the possibilities
(i) to study the elementary neutron amplitude,
(ii) to study the meson-nucleon interaction,
(iii) to study  medium effects, and 
(iv) to study nuclear structure.

\begin{figure}[h!]
 \plotinsert{-.2cm}{7cm}{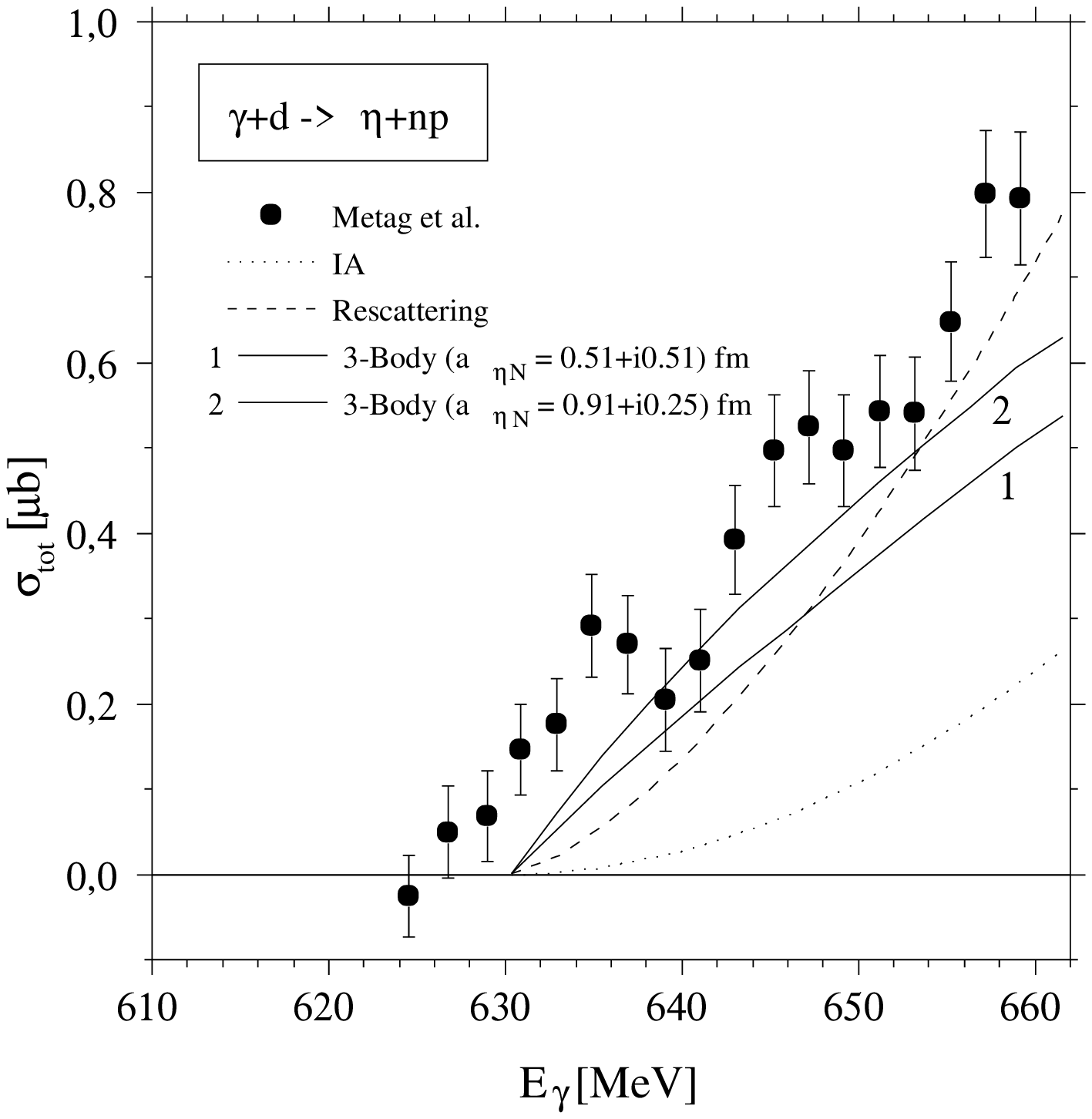}{-.7cm}
\mycaption{Total cross section for the reaction $\gamma d\to\eta np$.
Notation of the curves: dotted: impulse approximation (IA), dashed:
inclusion of first-order $\eta N$- and $NN$-rescattering, solid:
full three-body calculation.
The results obtained within the three-body approach with different
sets of $\eta NN^*$ and $\pi NN^*$ couplings are presented as the
solid curves ``1'' and ``2''.
The inclusive $\gamma d\to\eta X$ data are taken from Metag et al.\ 
(in preparation).}{-.3cm}
\label{fig11}
\end{figure}
\begin{figure}[h!]
 \plotinsert{-.2cm}{10cm}{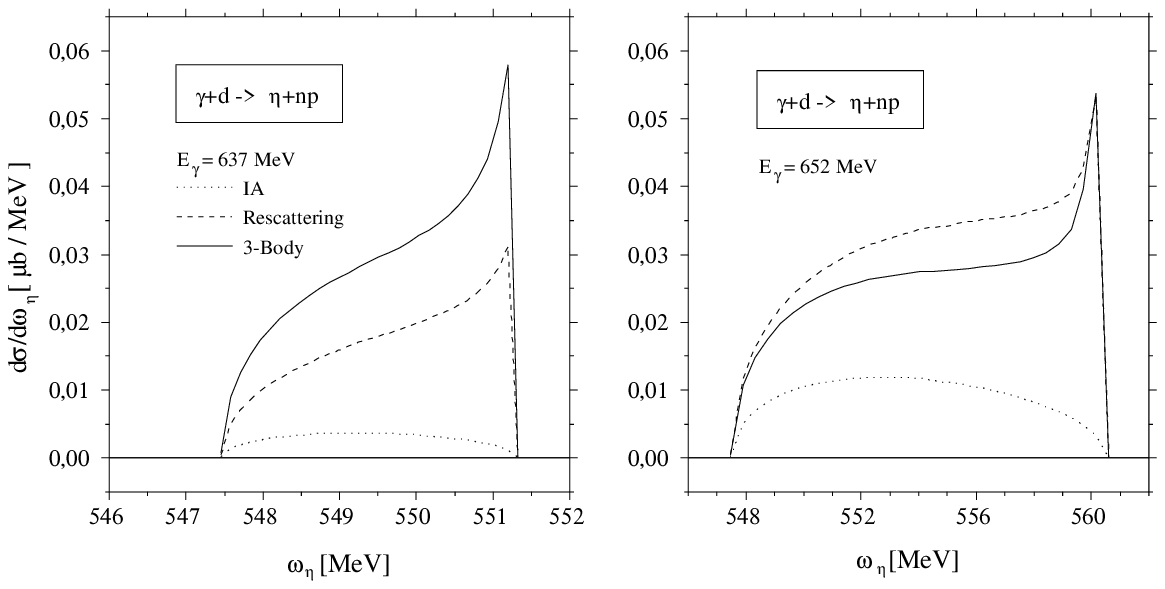}{-.7cm}
\mycaption{The $\eta$-meson spectrum for the reaction $\gamma d\to\eta np$ 
versus the total $\eta$ c.m.\ energy $\omega_\eta$ for two lab photon energies.
Notation of the curves as in Fig.~\ref{fig11}}{-.5cm}
\label{fig10}
\end{figure}

As an example of recent interest, I will present some very new results 
on incoherent $\eta$-meson production on deuterium near threshold. It turns 
out that the simple impulse approximation fails badly to describe the 
experimental data. This is seen very clearly in Fig.~\ref{fig11} where the 
total cross section is shown. The strong enhancement by final state 
interaction (FSI) is most evident. In fact, it is so strong that the first 
order rescattering contribution underestimates FSI considerably close to 
threshold, thus making a three-body calculation necessary which leads to a 
much improved description, although not completely satisfactory. But in 
this calculation, FSI has been included in $S$-waves only. A more detailed
test would be provided by a measurement of the meson spectrum shown in 
Fig.~\ref{fig10} which exhibits very clearly the signature of the virtual 
$^1S_0$-resonance of $NN$-scattering near threshold.

\section{Concluding remarks}
These examples should suffice for illustrating that the field of photo- 
and electronuclear physics has seen quite an impressive development both 
in breadth and depth over the past 40 years. This was made largely possible 
by the advent of new experimental tools and techniques and in theory by new 
ideas and the extraordinary progress in computing power. As a result, we 
have gained much deeper insights into the structure of hadrons, but still 
quite a few problems remain unresolved, in particular in the high energy 
regime, where a transition to quark-gluon dynamics is expected.

At present, the main research in this field is focused on the following topics:
\begin{itemize}
\item
short range structure of nuclei, in particular, the nature of 
2-body correlations,
\item
the size of relativistic effects in nuclear wave functions and operators,
\item
the role of subnuclear d.o.f.\ in terms of nuclear IC 
and meson exchange operators and their relation to the 
underlying quark-gluon d.o.f.\ of QCD,
\item
study of polarization observables,
\item
meson production on light nuclei. 
\end{itemize}

\vfill\eject
\end{document}